\documentclass[fleqn,usenatbib]{mnras}

\usepackage{newtxtext,newtxmath}

\usepackage[T1]{fontenc}

\DeclareRobustCommand{\VAN}[3]{#2}
\let\VANthebibliography\thebibliography
\def\thebibliography{\DeclareRobustCommand{\VAN}[3]{##3}\VANthebibliography}

\usepackage{graphicx}	
\usepackage{amsmath}	

\usepackage{booktabs}   

\defcitealias{DOGsample}{N19}
\defcitealias{Bussmann2012}{B12}
\defcitealias{Melbourne2012}{M12}

\title[DOGs in the radio]{Radio emission from dust-obscured galaxies}

\author[K. \'E. Gab\'anyi et al.]{
Krisztina \'E. Gab\'anyi,$^{1,2,3}$\thanks{E-mail: gabanyi@konkoly.hu} 
S\'andor Frey,$^{2,4}$ and
Krisztina Perger$^{2}$
\\
$^{1}$Department of Astronomy, E\"otv\"os Lor\'and University, P\'azm\'any P\'eter s\'et\'any 1/A, H-1117 Budapest, Hungary\\
$^{2}$Konkoly Observatory, ELKH Research Centre for Astronomy and Earth Sciences, Konkoly Thege Mikl\'os \'ut 15-17,
H-1121 Budapest, Hungary\\
$^{3}$Extragalactic Astrophysics Research Group, E\"otv\"os Lor\'and University, P\'azm\'any P\'eter s\'et\'any 1/A, H-1117 Budapest, Hungary\\
$^{4}$Institute of Physics, ELTE E\"otv\"os Lor\'and University, P\'azm\'any P\'eter s\'et\'any 1/A, H-1117 Budapest, Hungary
}

\date{Accepted XXX. Received YYY; in original form ZZZ}

\pubyear{2021}

\begin{document}
\label{firstpage}
\pagerange{\pageref{firstpage}--\pageref{lastpage}}
\maketitle

\begin{abstract}
The coevolution of galaxies and their central supermassive black holes is a subject of intense research. A class of objects, the dust-obscured galaxies (DOGs) are particularly interesting in this respect as they are thought to represent a short evolutionary phase when violent star formation activity in the host galaxy may coexist with matter accretion onto the black hole powering the active nucleus. Here we investigate different types of DOGs classified by their mid-infrared spectral energy distributions to reveal whether they can be distinguished by their arcsec-scale radio properties. Radio emission is unaffected by dust obscuration and may originate from both star formation and an active nucleus. We analyse a large sample of 661 DOGs complied from the literature and find that only a small fraction of them ($\sim 2$ per cent) are detected with flux densities exceeding $\sim 1$\,mJy in the Faint Images of the Radio Sky at Twenty-Centimeters (FIRST) survey. These radio-detected objects are almost exclusively `power-law' DOGs. Stacking analysis of the FIRST image cutouts centred on the positions of individually radio-undetected sources suggests that weak radio emission is present in `power-law' DOGs. On the other hand, radio emission from `bump' DOGs is only marginally detected in the median-stacked FIRST image.
\end{abstract}




\section{Introduction}

Enhanced star formation and activity of the central supermassive black hole in galaxies as an active galactic nucleus (AGN) are both relatively short phases in the lifetime of a galaxy. The two phenomena are linked together, and the peak of both the star formation \citep[e.g.][]{Heavens2004} and the enhanced nuclear activity occurred at a similar cosmological epoch, at redshift $z\sim 2-3$ \citep[e.g.][]{1995AJ....110...68S,2000MNRAS.311..576K}.

The large amount of interstellar gas and dust present in star forming galaxies and obscured AGN hinder their detection in the optical bands. Methods employing observations at other wavelengths, e.g., hard X-ray detection of obscured AGN, or various colour selection techniques are used. In the sub-mm and infrared regimes, the emission of colder and hotter dust can be detected. The dust can be heated by star formation and/or the AGN in these systems, and the determination of the heating source can often be challenging, especially at larger redshifts \citep[e.g.,][and references therein]{Farrah2017}. Among the sub-mm galaxies (SMG) originally selected as intensely star-forming galaxies, several are found to contain obscured AGN or their spectral energy distribution (SED) can be equally well fitted with star-forming and AGN templates \citep[e.g.,][and references therein]{SMG_new}. In ultraluminous infrared galaxies \citep[ULIRGs;][]{ulirg}, the infrared luminosities exceeding $10^{12} L_\odot$ can originate from dust heating caused by highly intensive star formation episodes and/or accretion to the central supermassive black hole of the galaxy.

\cite{Dey2008} identified dust-obscured galaxies (DOGs) in the NOAO Deep Wide-Field Survey Bo\"otes field, which were missed in previous optical surveys. These objects are characterized by  $S_{24\mu\text{m}}>0.3$\,mJy and $S_{24\mu\text{m}}/S_R>1000$ , where $S_{24\mu\text{m}}$ and $S_R$ stand for the $24$-$\mu$m and the {\it R}-band flux densities, respectively and exhibit evidences for both star formation and AGN.

The DOGs can be divided into two groups based upon their mid-infrared SEDs. The `bump' (B) DOGs show a peak (or bump) at the rest-frame wavelength of $1.6\,\mu$m, while the SEDs of `power-law' (PL) DOGs are dominated by a power law in the mid-infrared regime \citep{Dey2008}. The $1.6$-$\mu$m bump in the SED arises due to the thermal emission of late-type stars, and is enhanced by the minimum opacity of H$^-$ at $\sim 1.6\,\mu$m \citep[][and references therein]{bump_selection2}. Thus, the shape of the SED, and the infrared spectroscopic characteristics of B DOGs indicate that their bolometric luminosities are dominated by star formation. While the existence of the $1.6$-$\mu$m bump in the SED implies that AGN continuum does not contribute significantly to the near-IR  emission, it does not rule out completely the existence of an AGN in the system \citep{bump_selection2}. On the other hand, AGN activity is thought to dominate the emission seen in PL DOGs.

The AGN contribution is more established in an extreme subsample of DOGs, the so-called hot DOG sources. They were classified in a sample of luminous infrared galaxies detected by the {\it Wide-field Infrared Survey Explorer} space telescope \citep[{\it WISE},][]{wise}, and found to fulfil the DOG criteria. These objects are hardly detected in the two shorter wavelength filters of {\it WISE} in $W1$ and $W2$ bands, corresponding to $3.4\,\mu$m and $4.6\,\mu$m, but they are bright sources in the $W3$ and $W4$ bands, corresponding to $12\,\mu$m and $22\,\mu$m, respectively \citep{Eisenhardt2012, Wu2012}. Based upon their infrared luminosities, hot DOGs belong to the hyperluminous infrared galaxies ($L_\mathrm{IR}>10^{13} \mathrm{L}_\odot$). \cite{Tsai2015_WISE} investigated $20$ even more extreme hot DOGs, dubbed as extremely luminous infrared galaxies (ELIRGs) with $L_\mathrm{IR}>10^{14} \mathrm{L}_\odot$. They found that these objects most probably contain highly obscured AGNs responsible for the high infrared luminosities. \cite{Farrah2017} studied a dozen hot DOGs, the most luminous objects at $z\sim2$, and found that obscured AGN can explain the SED and the infrared colors. However, star formation rate (in the order of magnitude) of $100\,\mathrm{M}_\odot \mathrm{\,yr}^{-1}$ are still possible in some sources. \cite{Assef2016} identified hot DOGs with optical/ultraviolet emission exceeding that expected from star formation, which can arise from leaked emission from obscured AGN (or from a second unobscured AGN), as well as a young starburst.

Other infrared-bright galaxies with luminosities in the regime of LIRGs were found among the Lyman-$\alpha$ emitters by \cite{bridge2013}. They are more luminous and contain warmer dust than DOGs and SMGs. The rarity of these objects indicates that they represent a short-lived phase between the starburst and the optical quasar stage. A similar, transient AGN-dominated phase was used to describe the characteristics of the {\it WISE} and radio-selected luminous, high-redshift, dusty galaxy sample studied by \cite{Jones2015}.

\citet{Bussmann2012} (hereafter B12) and \citet{Melbourne2012} (hereafter M12) proposed an evolutionary sequence in which an intense star-formation period induced by a gas-rich major merger leads to an observable SMG or B DOG. Later, as the star formation slows down and black-hole growth picks up, a PL DOG can be observed. \citetalias{Bussmann2012} studied a subsample of DOGs detected by \citet{Dey2008}. They focused on objects with spectroscopic redshifts above $1.4$ and found tentative evidence supporting the evolutionary scenario where SMGs evolve first into B DOGs and then into PL DOGs.

\citet{DOGsample} (hereafter N19) created a sample of $571$ infrared-bright DOGs using optical and infrared data obtained with the Subaru Hyper Suprime-Cam \citep[HSC; ][]{HSC_DR1} survey, the VISTA Kilo-degree Infrared Galaxy survey \citep[VIKING DR2; ][]{Viking_DR2}, and the {\em Wide-field Infrared Survey Explorer} all-sky survey \citep[AllWISE; ][]{allwise}. They found that the $(g-z)_\text{AB}$ colour of PL DOGs is bluer than that of the B DOGs. This is consistent with the evolutionary scenario where the star formation dominated phase (observable as B DOGs) is followed by the AGN-dominated phase (observable as PL DOGs). 

The radio emission in a galaxy may either be linked to star-formation or AGN activity. While optical, UV, and X-ray radiation is absorbed by the the dense dusty environment, radio measurements are largely unaffected by it. Thus, they hold the potential to infer the presence of an active (jetted) nucleus even in heavily obscured host galaxies if we have an independent estimate of the rate of star formation \citep[e.g., ][and references therein]{2021review}. Radio emission of hot DOGs was studied by observations with the Karl G. Jansky Very Large Array (VLA) and Atacama Large Millimeter/Submillimeter Array (ALMA) and at higher resolution with very long baseline interferometry \citep{hotDOG_VLA,hotDOG_ALMA, hotDOG_Frey}. These observations showed that star-formation and nuclear activity co-exist and revealed objects with newly triggered, young jets still embedded in the dusty material of their host galaxies. 
To study the radio properties of the different types of DOG sources, we cross-matched B and PL DOG samples with the Faint Images of the Radio Sky at Twenty-Centimeters \citep[FIRST;][]{FIRST_Helfand} survey catalogue. We investigated individually those few sources having FIRST detections, and performed stacking analysis for the FIRST-undetected DOGs.

In the following, we assume a flat $\Lambda$CDM cosmological model with $H_0=70\,\text{km\,s}^{-1}\text{Mpc}^{-1}$, $\Omega_\text{m}=0.27$, and $\Omega_\Lambda=0.73$.

\section{Sample selection and data}

\begin{table*}
    \caption{Details of the stacked DOG samples. The names, right ascension and declination in degrees, and redshifts are given in columns 1, 2, 3, and 4, respectively. In column 5, the redshift flag indicates whether it is a spectroscopic (0) or photometric (1) redshift. The DOG classification is indicated in column 6, here the blue-excess DOG classification of \citetalias{DOGsample} is given as `BluDOG'. In the last column, we indicate from which sample the source is drawn, N2019, B2012, and M2012 stands for \citetalias{DOGsample}, \citetalias{Bussmann2012}, and \citetalias{Melbourne2012}, respectively. The full table is available in electronic format online as Supplementary material.}
    \centering
    \begin{tabular}{ccccccc}
    \hline
    Name & RA ($\degr$) & Dec ($\degr$) & Redshift & Redshift flag & Class. & Ref. \\
    \hline
    HSC J021647.48$-$041334.6 & $34.19785$ & $-4.22628$ &  $0.96$ & 1& U &        N2019\\
HSC J021656.62$-$051005.4 & $34.23591$ &$-5.16816$ &  $1.11$ & 1 & PL &        N2019\\
HSC J021718.52$-$034350.1 & $34.32716$ & $-3.73057$ &  $0.39$ & 1 & PL        & N2019 \\
HSC J021729.07$-$041937.6 & $34.37111$ & $-4.32712$ &  $0.95$ & 1 & PL  &       N2019 \\
HSC J021742.81$-$034531.0 & $34.42836$ & $-3.75861$ &   $0.64$ & 1 & PL &        N2019 \\
HSC J021749.02$-$052306.7 & $34.45423$ & $-5.38521$ &  $1.47$ & 1 & PL &        N2019 \\
HSC J021754.45$-$043015.7 & $34.47687$ & $-4.50436$ &  $1.13$ & 1 & PL &        N2019 \\
    \end{tabular}
    \label{tab:stacked_DOGs}
\end{table*}

\citetalias{DOGsample} used optical observations from the HSC survey, infrared data from the VIKING DR2 and the AllWISE catalogues to create a sample of infrared-bright DOGs. They first excluded measurements possibly affected by bad pixels, cosmic rays, neighbouring bright sources, saturated pixels, or other significant noises. Moreover, they excluded objects from the optical and near-infrared data sets which had signal-to-noise ratio (SNR) less than 5 in any of the used bands. They also excluded the AllWISE sources with $\mathrm{SNR}<3$ in band $W4$ and those flagged as extended. After obtaining these clean samples, they first cross-matched the optical and near-infrared samples and employed a colour cut of $(i-K_S)_\text{AB}\geqslant 1.2$ following \citet{Toba2015}. Then they cross-matched the result with the AllWISE clean sample and used the DOG selection criterion of $(i-[22])_\text{AB}\geqslant 7.0$. Thus they obtained a catalog of $571$ DOGs.

\citetalias{DOGsample} assumed a power-law for {\it WISE} $W2$, $W3$, and $W4$ data points and calculated the expected $K_S$ band flux density. If the observed $K_S$ band flux density was larger than $3$ times the value expected from the power-law fit, they classified the source as B DOG.  This way, they had $51$ B DOGs and $257$ PL DOGs in their sample. Only the sources with detections in all of the three WISE bands ($W2$, $W3$, and $W4$) with $\mathrm{SNR} \geqslant 2$ were classified. The remaining $263$ sources are the unclassified (U) DOGs.

\citetalias{DOGsample} applied the MIZUKI code of \cite{mizuki_code1} to determine the photometric redshifts of their sources. Following the criteria of \cite{Toba2017}, they defined the reliable photometric redshift value if the reduced $\chi^2$ of template fitting is smaller than 1.5, and the relative error of the derived redshift is smaller than $5$ per cent. In their sample, $152$ DOGs have reliable redshift estimate.\footnote{Unfortunately, they do not indicate in their published data whether the photometric redshift of a given source is reliable.} They also found, however, that the averages and standard deviations of all the photometric redshifts and the reliable ones are not significantly different.

The DOG sample of \citetalias{Bussmann2012} is based on the {\it Spitzer}-selected DOG catalogue of \cite{Dey2008}. \cite{Dey2008} selected sources in the NDWFS Bo\"otes Field satisfying the following criteria: $R-[24]>14$ (in Vega magintudes), and $S_{24\mu\mathrm{m}}>0.3$\,mJy. \citetalias{Bussmann2012} focused on those having known redshifts above $1.4$ ($90$ sources). In this sample, there are $31$ B DOGs and $59$ PL DOGs\footnote{In the original publication of \citetalias{Bussmann2012}, the numbers in the different groups are mistyped.}. We complemented this list with $27$ additional DOGs ($16$ B DOGs and $11$ PL DOGs) reported in \citetalias{Melbourne2012} which are originally from the same DOG catalogue of \cite{Dey2008}, but were not investigated by \citetalias{Bussmann2012} because of either unknown (for them) or too low redshift values. All $117$ DOG sources from the combined \citetalias{Bussmann2012} and \citetalias{Melbourne2012} samples have spectroscopic redshifts.

\citetalias{DOGsample}, following \cite{Toba2015}, used an optical--near-IR colour cut in their selection of DOG sources, $(i-K_S)_\text{AB}\geqslant 1.2$. According to \cite{Toba2015}, all the DOG sources of \citetalias{Bussmann2012} also satisfy this criteria. However, the DOG sources studied by \citetalias{DOGsample} are brighter in the $22$-$\mu$m band (with $S_{22\mu\mathrm{m}}\sim 4-10$\,mJy) than those in the \citetalias{Bussmann2012} sample in the $24\mu$m band, therefore \citetalias{DOGsample} named their sample as infrared-bright DOGs. As \citetalias{DOGsample} states, this can result in losing some relatively blue DOGs from their sample.

All the $661$ DOG sources of \citetalias{DOGsample}, \citetalias{Bussmann2012}, and \citetalias{Melbourne2012} are within the coverage of the FIRST survey \citep{FIRST_Helfand}. We found that $19$ DOGs of the \citetalias{DOGsample} sample and $6$ sources from the \citetalias{Bussmann2012} and \citetalias{Melbourne2012} samples have detected radio counterparts within $1\farcs5$ of their positions in the latest FIRST catalog \citep{FIRST_Helfand}. The angular separation limit was chosen following \citet{Ivezic2002}. These radio-detected objects were left out from our subsequent stacking analysis. 

We also disregarded sources with positions falling close (within $2\farcm25$) to the edge of the FIRST survey coverage area ($30$ sources from the \citetalias{DOGsample} sample). Since we used image cutouts of $4\farcm5 \times 4\farcm5$, that meant to exclude objects with incomplete cutouts. Additionally, we only included those sources for which \citetalias{DOGsample} listed photometric redshift estimates. 
Thus, we finally used the positions of $224$ PL, $48$ B, and $245$ U DOGs of the \citetalias{DOGsample} sample and additional $64$ PL and $47$ B DOGs of the \citetalias{Bussmann2012} and \citetalias{Melbourne2012} samples in the stacking analysis. Details of these objects are summarized in Table \ref{tab:stacked_DOGs}. Their non-detection in FIRST implies that their $1.4$-GHz flux densities are typically $\la 1$\,mJy. The $4\farcm5 \times 4\farcm5$ cutout images centred on the DOG positions were downloaded from the FIRST archive\footnote{\url{http://sundog.stsci.edu}} in Flexible Image Transport System \citep[FITS;][]{fits} format.

\section{Stacking analysis}  

\begin{table*}
	\centering
	\caption{Results of the stacking of the different groups of FIRST-undetected DOGs, their designations are given in the first column. Above and below the line, we list the stacking results for the \citetalias{DOGsample} sample, and for the mixed sample \citepalias{Bussmann2012,Melbourne2012}, respectively. We list the number of stacked images, the average and the standard deviations of the redshifts, in columns 2 and 3. In columns 4, 5, and 6, the intensity values of the central pixel, the image rms noise levels ($1 \sigma$), and the signal-to-noise ratios are listed, respectively. In the last three columns, we give the average peak, rms noise level and the corresponding signal-to-noise ratios for the median-stacked images of the randomly chosen cutouts (see the text for details). The pixel values were calculated by taking the bias described in \citet{FIRST_stack_2007} into account.}
	\label{tab:stacking}
	\begin{tabular}{lcccccccc} 
		\hline
		\multicolumn{6}{c}{Stacked DOG cutouts} & \multicolumn{3}{c}{Stacked empty-field cutouts} \\
		\cmidrule(lr){3-6} \cmidrule(lr){7-9}
		ID & No. & Average & Central pixel value & rms noise level & SNR & Average peak & Average rms noise level & SNR \\ 
		 & & redshift & $\mu$Jy\,beam$^{-1}$ & $\mu$Jy\,beam$^{-1}$ & & $\mu$Jy\,beam$^{-1}$ & $\mu$Jy\,beam$^{-1}$ & \\
		\hline
		PL & $224$ & $1.15 \pm 0.41$ & $181.2$ & $17.8$ & $10.2$ & $67.8$  & $16.5$ & $4.1$ \\ 
		B & $48$ & $0.99\pm 0.39$ & $134.0$ & $36.6$ & $3.7$ & $148.0$ & $36.0$ & $4.1$ \\ %
		U & $245$ & $1.06\pm0.29$ & $12.9$ & $15.9$ & $0.8$ & $63.3$ & $15.6$ & $4.1$ \\ 
		\hline
		PL & $288$ & $1.37 \pm 0.61$ & $158.6$ & $15.6$ & $10.2$ & $58.8$ & $14.4$ & $4.1$\\
		B & $95$ & $1.34\pm0.62$ & $100.3$ & $26.3$ & $3.8$ & $103.8$ & $25.5$ & $4.1$ \\
	\end{tabular}
\end{table*}

\begin{figure*}
\begin{minipage}{0.33\textwidth}
\centering
	\includegraphics[width=\columnwidth]{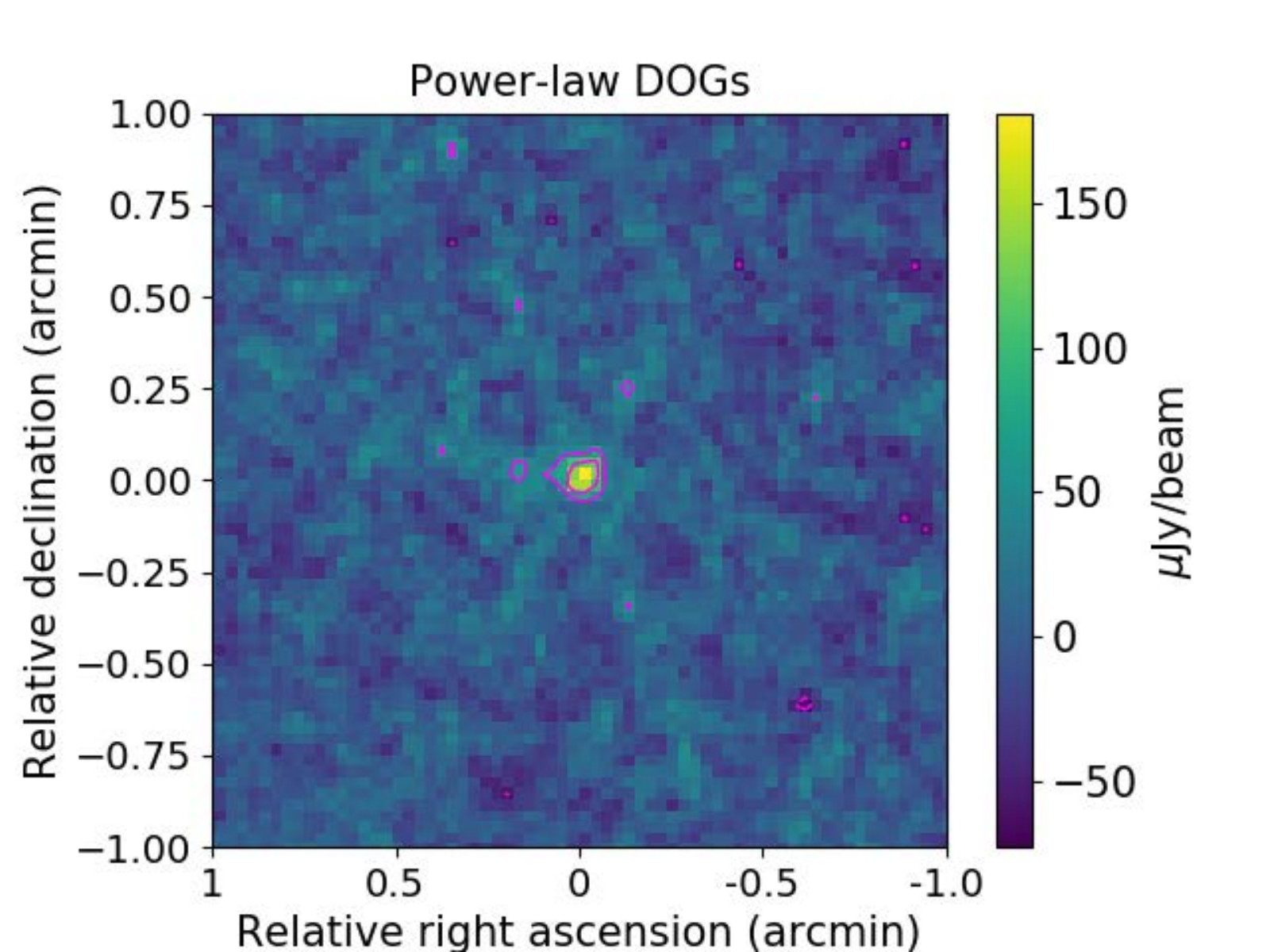}
\end{minipage}
\hfill
\begin{minipage}{0.33\textwidth}
\centering
	\includegraphics[width=\columnwidth]{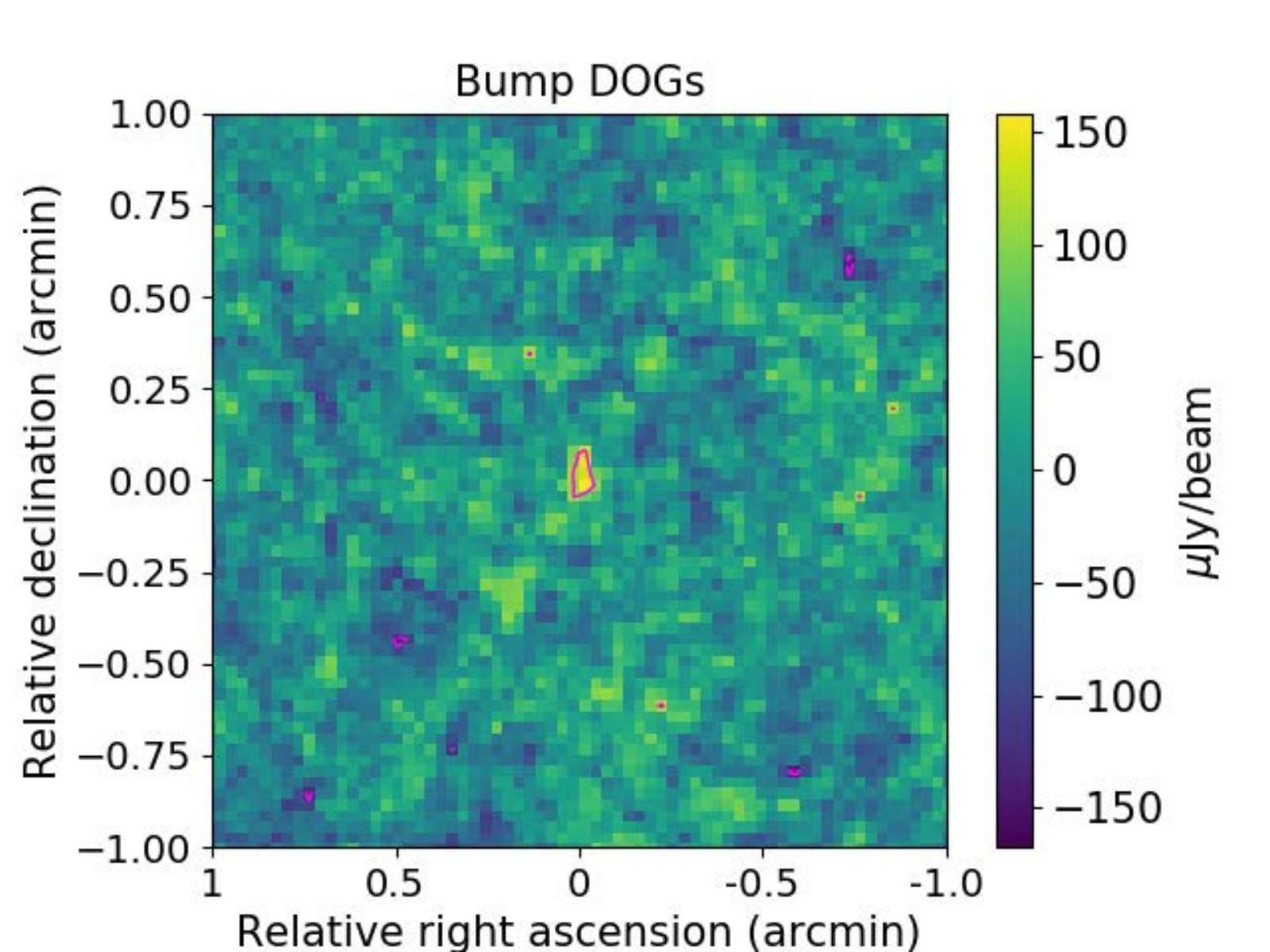}
\end{minipage}
\hfill
\begin{minipage}{0.33\textwidth}
\centering
	\includegraphics[width=\columnwidth]{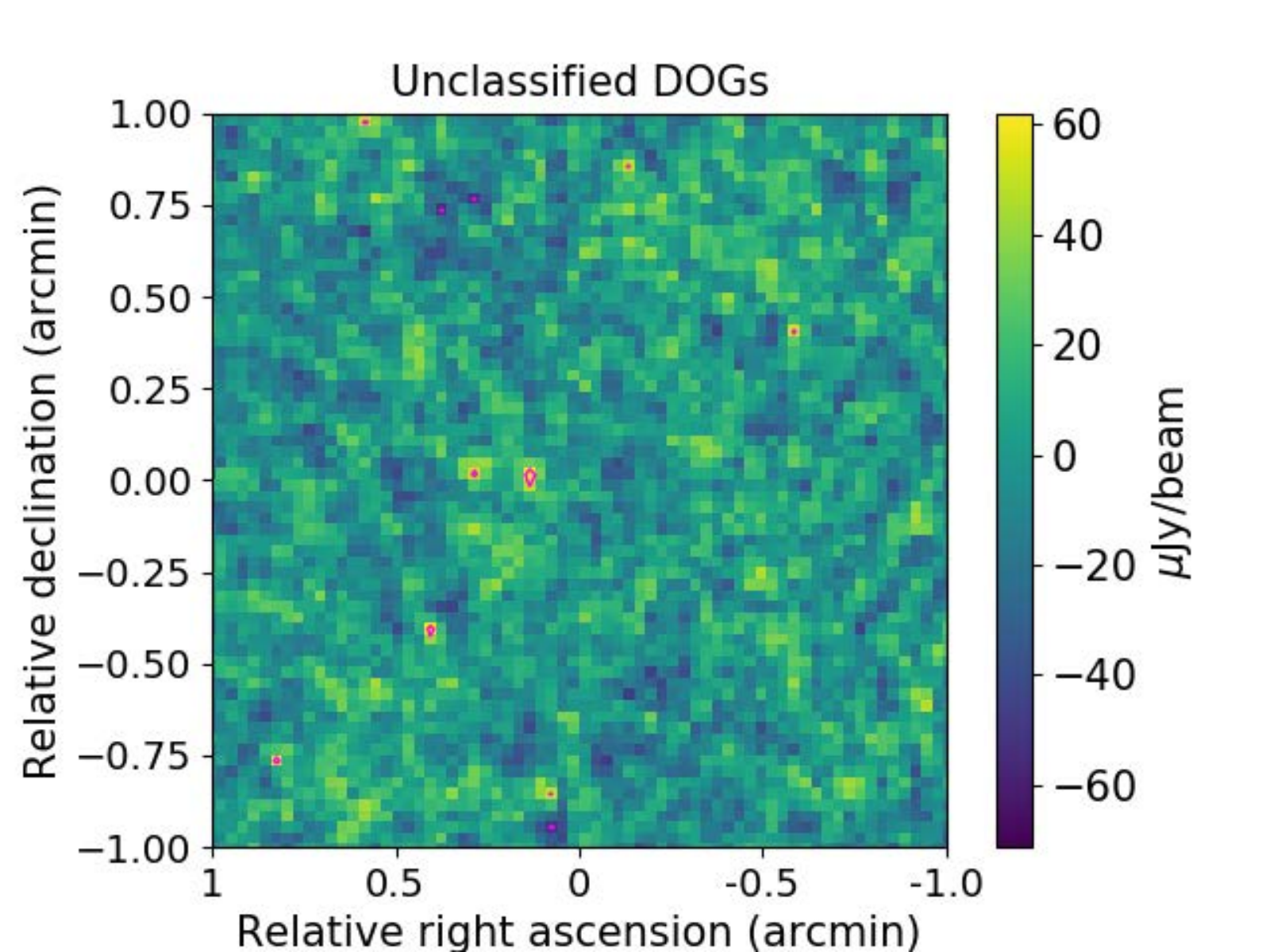}
\end{minipage}
    \caption{The stacked FIRST images of the three groups of undetected DOG source positions taken from the \citetalias{DOGsample} sample.}
    \label{fig:stack}
\end{figure*}

The method of stacking is often used to reveal the faint emission of a group of objects having flux densities below the detection threshold of a given survey. Median stacking, i.e. calculating the median intensity for each pixel in the stack of images, has the advantage to exclude the outlier pixel values. Median stacking of FIRST cutout images were used to look for the faint radio emission of e.g., optically identified quasars \citep{FIRST_stack_2007}, low-luminosity AGN \citep{FIRST_stack_SF}, radio-quiet quasars \citep{FIRST_stack_2008,FIRST_stack_2018}, special class of quiescent galaxies with ionized gas emission features \citep{FIRST_stack_geyser}, and high-redshift AGN \citep{Perger_stack}.

To investigate the $1.4$-GHz radio emission of faint ($\la 1$\,mJy) DOG sources and to compare the radio characteristics of the different types of DOGs, we stacked the downloaded FIRST cutout images and calculated the pixel-by-pixel median values for each of the DOG subclasses. Following the prescription of \citet{FIRST_stack_2007}, we multiplied the pixel values of the stacked images by $1.4$ to correct for the so-called snapshot bias. We first focused on the \citetalias{DOGsample} sources, and stacked only the positions of these objects. The resulting images are shown in Fig.~\ref{fig:stack}. The intensity values of the central pixels, the root-mean-square (rms) noise levels, and the signal-to-noise ratios (SNR) obtained for the three different groups of sources are listed in the upper part of Table~\ref{tab:stacking}. In the case of the PL DOGs, the central pixel value of the median-stacked images is ten times the rms noise level of the image. On the other hand, in the case of B and U DOGS, the central pixel values are only $3.7$ and $0.8$ of the respective median stacked image noise levels, thus far below the SNR obtained for PL DOGs.

To assess the significance of these values, we also stacked FIRST image cutouts centred at random positions. We downloaded $555$ such FIRST image cutouts, and randomly selected from this pool the same number of images used for stacking the DOG image cutouts. We then followed the same steps to obtain the median stacked images. We repeated this procedure $50$ times, and calculated the average of maximum pixel values, rms noise levels and corresponding SNRs. The results are given in Table~\ref{tab:stacking}. The brightest pixel in the stacked image can have $\mathrm{SNR} \approx 4$ even when FIRST image cutouts at random positions are stacked, and thus no radio emission is expected in the median image. The noise level of the median images decreases with $\sqrt{N}$, where $N$ is the number of stacked images, as expected. We note that the pixels with the highest intensity in these images do not fall at the image centres but were located at various random positions. Thus, these values indeed characterize the random noise distribution of FIRST cutout images. Similar results, SNRs between $4-5$, were achieved by \cite{Perger_stack} when median stacking of FIRST images around fake positions were performed.

We detected radio emission in the stacked PL DOG positions of the \citetalias{DOGsample} sample with $\mathrm{SNR} \sim 10$. This clearly exceeds the SNR obtained from stacking image cutouts at random positions. We used the IMFIT task of the NRAO Astronomical Image Processing System \citep[AIPS; ][]{aips} to fit a Gaussian brightness distribution model to the PL DOGs stacked median image. The resulting flux density is $(328\pm 60)\,\mu$Jy, the peak intensity is $(119 \pm 17)\,\mu\text{Jy\,beam}^{-1}$, the major axis size (full width at half-maximum, FWHM) is $13\farcs0 \pm 1\farcs8$, and a minor axis FWHM is $7\farcs3\pm1\farcs0$ with the major axis oriented at the position angle of $87\degr \pm9\degr$ (measured from north through east). The resulting ellipticity does not have any physical meaning, since the stacked sources are oriented at random position angles. The asymmetries are related to beam effects \citep{FIRST_stack_2007}. The obtained axis sizes are larger than the size expected for a point source observed in FIRST \citep{FIRST_stack_2007}. This could arise if the faint radio sources do not exactly coincide with the optical (and infrared) positions and/or the radio emission originates from a slightly extended region.

There is no detection in the stacked images centred on U DOGs  positions. In the case of the B DOGs, the $\text{SNR}$ of the central pixel is $\sim3.7$. Since similar or even slightly higher $\text{SNR}$ values can be obtained when stacking the image cutouts at random positions (Table~\ref{tab:stacking}), this value cannot be regarded as an indication of a convincing detection. However, the brightest pixels of the median-stacked image of B DOGs appear in the central part of the image where radio emission is expected if present, while a random brightness peak could show up at any position of the stacked image. It is indeed the case for the stacked image cutouts centred at random positions. Therefore, based on both the SNR and the peak location, we consider B DOGs marginally detected in their median-stacked FIRST image. 

The number of stacked B DOG images is much lower than for the PL and U DOGs, therefore the achieved noise level in the stacked median image is around twice as high as for the other two groups. To reduce the rms noise level of the stacked image of B DOGs, we repeated the stacking procedure with including the positions of the DOG sources from the \citetalias{Bussmann2012} and \citetalias{Melbourne2012} samples. The values obtained for the central peak intensity, the rms noise levels, and the SNR are listed in the lower part of Table~\ref{tab:stacking}. The SNR of the central pixel values did not change profoundly. Radio emission from the median-stacked image of PL DOGs is detected at $\sim10$ times the noise level, while for the B DOGs the SNR of the central pixel of the stacked median image remains below $4$.

\section{Detected sources}

\begin{table*}
\caption{Details of the FIRST-detected DOGs. In the upper part, the FIRST-detected sources from \citetalias{DOGsample} are listed, here only photometric redshifts ($z_\text{ph}$) are available. In the lower part, the FIRST-detected sources from \citetalias{Bussmann2012} and \citetalias{Melbourne2012} are listed, they have spectroscopic redshifts. The name and classification of each source is given in columns 1 and 2, respectively. The redshift (either photometric or spectroscopic), and the luminosity distance ($D_\text{L}$) is given in columns 3 and 4, respectively. We list the sidelobe probability, peak intensity, integrated flux density, and rms noise level \citep[all from the FIRST catalog of ][]{FIRST_Helfand} in columns 5, 6, 7, and 8, respectively. In the last two columns, the compactness parameter (i.e. the ratio of the peak intensity to the integrated flux density) and the $1.4$-GHz radio power are given.   
\label{tab:FIRST_det}}
\begin{tabular}{lccccccccc}
\hline
ID & Class & $z_\text{ph}$ / $z$ & $D_\text{L}$ & $p_{\text{s}}$ & $S_\text{peak}$ & $S_\text{int}$ & rms & Comp. & $P_{1.4\text{GHz}}$ \\
 & & & Mpc &  & mJy\,beam$^{-1}$ & mJy & mJy & & $10^{22}$ W\,Hz$^{-1}$\\
\hline
  HSC J083713.56$+$021707.1 & PL & $1.05$ & $7151.9$ & $0.014$ & $1.23$ & $1.02$ & $0.141$ & $1.21$ & $32$\\ 
  HSC J085552.54$+$014553.2 & PL & $1.50$ & $11161.1$ & $0.014$ & $2.44$ & $2.07$ & $0.154$ & $1.18$ & $130$\\ 
  HSC J091128.32$-$011812.2 & PL & $1.53$ & $11445$ & $0.014$ & $70.32$ & $72.62$ & $0.145$ & $0.97$ & $4720$\\ 
  HSC J091218.78$-$000401.3 & PL & $1.38$ & $10063.1$ & $0.014$ & $17.89$ & $19.3$ & $0.14$ & $0.93$ & $1030$ \\ 
  HSC J091854.50$+$015046.6$^*$ & PL & $1.12$ & $7752.6$ & $0.528$ & $1.13$ & $1.14$ & $0.152$ & $0.99$ & $41$ \\ 
  HSC J140256.23$-$000347.7 & PL & $1.55$ & $11631.7$ & $0.058$ & $1.15$ & $0.81$ & $0.133$ & $1.42$ & $54$ \\ 
  HSC J140444.81$+$002451.1 & B & $0.05$ & $222.5$ & $0.014$ & $10.95$ & $11.44$ & $0.151$ & $0.96$ & $0.7$ \\ 
  HSC J140638.20$+$010254.6 & PL & $0.236^\text{a}$ & $1188.4$ & $0.014$ & $8.24$ & $9.19$ & $0.154$ & $0.90$ & $13.2$\\ 
  HSC J140738.47$+$002731.4 & PL & $0.54$ & $3142.9$ & $0.014$ & $1.55$ & $0.89$ & $0.15$ & $1.74$ & $7.2$\\ 
  HSC J141546.89$-$011451.7$^*$ & PL & $1.06$ & $7237.1$ & $0.452$ & $1.05$ & $3.01$ & $0.154$ & $0.35$ & $96$ \\ 
  HSC J141955.59$-$003449.1 & PL & $1.95$ & $15466.4$ & $0.014$ & $16.89$ & $17.48$ & $0.145$ & $0.97$ & $1780$ \\ 
  HSC J142207.90$-$002408.5 & U & $0.99$ & $6645.2$ & $0.014$ & $4.11$ & $3.62$ & $0.152$ & $1.13$ & $101$ \\ 
  HSC J142332.74$-$012526.8 & U & $0.52$ & $3002.2$ & $0.014$ & $1.45$ & $1.66$ & $0.15$ & $0.87$ & $12$\\ 
  HSC J142435.43$+$005947.8$^*$ & PL & $1.22$ & $8627.3$ & $0.188$ & $1.19$ & $1.34$ & $0.155$ & $0.89$ & $56$\\ 
  HSC J143004.57$+$011653.4 & U & $0.86$ & $5575.6$ & $0.014$ & $2.85$ & $2.9$ & $0.137$ & $0.98$ & $61$ \\ 
  HSC J144557.52$-$002847.0 & PL & -- & -- & $0.014$ & $2.92$ & $2.81$ & $0.172$ & $1.04$ & -- \\ 
  HSC J144740.41$-$013524.0 & PL & $0.98$ & $6561.5$ & $0.014$ & $1.35$ & $1.78$ & $0.146$ & $0.76$ & $49$ \\ 
  HSC J144814.45$+$005302.3 & PL & $0.98$ & $6561.5$ & $0.098$ & $1.11$ & $0.71$ & $0.14$ & $1.56$ & $19$ \\ 
  HSC J145753.79$+$000508.5 & PL & $1.12$ & $7752.6$ & $0.014$ & $1.94$ & $1.93$ & $0.152$ & $1.00$ & $69$ \\ 
  \hline
  SST24 J142648.9$+$332927 & PL & $2.00$ & $15957.9$ & $0.014$ & $1.63$ & $1.42$ & $0.142$ & $1.15$ & $151$ \\ %
  SST24 J142842.9$+$342409 & PL & $2.18$ & $17746.6$ & $0.039$ & $1.25$ & $1.29$ & $0.132$ & $0.97$ & $161$ \\
  SST24 J143102.2$+$325152 & PL & $2.00$ & $15957.9$ & $0.014$ & $2.10$ & $1.57$ & $0.116$ & $1.34$ & $167$ \\
  SST J143430.6$+$342757 & PL & $1.24$ & $8804.4$ & $0.014$ & $1.37$ & $1.49$ & $0.134$ & $0.92$ & $65$ \\
  SST J143541.2$+$334228 & PL & $1.39$ & $10154.2$ & $0.014$ & $1.26$ & $1.46$ & $0.13$ & $0.86$ & $79$ \\
  SST24 J143644.2$+$350627 & PL & $1.95^\text{b}$ & $15957.9$ & $0.014$ & $4.25$ & $4.81$ & $0.142$ & $0.88$ & $522$ \\ %
  \hline
  \multicolumn{10}{l}{$^*$ The FIRST source can be a spurious detection caused by sidelobes of unrelated bright sources.} \\
  \multicolumn{10}{l}{$^\text{a}$ The used value is spectroscopic redshift from SDSS DR9 \citep{SDSS_DR9}. \citetalias{DOGsample} used $z_\text{ph}=0.68$.} \\ 
  \multicolumn{10}{l}{$^\text{b}$ A slightly different spectroscopic redshift, $z=1.84$ is given by \cite{2016MNRAS.455.1796H}.}
\end{tabular}
\end{table*}

The details of the FIRST-detected DOGs are summarized in the Table \ref{tab:FIRST_det}. The upper part displays the $19$ sources from the \citetalias{DOGsample} sample, the lower part the $6$ sources from the \citetalias{Bussmann2012} and \citetalias{Melbourne2012} samples.

According to \cite{FIRST_Helfand}, applying a cut for the sidelobe probability at $p_{\text{s}}<0.1$ in the FIRST survey eliminates most of the false detections caused by the sidelobes of unrelated bright sources in the vicinity. We checked the images of those sources where the FIRST catalogue gives $p_{\text{s}}>0.1$, HSC\,J091854.50$+$015046.6, HSC\,J141546.89$-$011451.7, and  HSC\,J142435.43$+$005947.8. Since we cannot exclude the possibility that they are sidelobes, we leave them out of further discussion.

Most of the FIRST-detected DOGs are PL DOGs ($12+6$), while one B DOG and three U DOGs also have radio counterparts. From the DOG sources of \citetalias{DOGsample}, $\sim5$ per cent of PL DOGs, $\sim2$ per cent of B DOGs, and $\sim1$ per cent of U DOGs were detected in FIRST. From the \citetalias{Bussmann2012} and \citetalias{Melbourne2012} sample, $\sim 8$ per cent of the PL DOGs have radio counterparts in FIRST and none of the B DOGs. The only radio-detected B DOG has the lowest photometric redshift estimate, much lower than the typical value of $z\sim1$.

The ratio of the peak intensity ($S_\text{peak}$) to the flux density ($S_\text{int}$) of the individual sources is close to $1$. This indicates that most of the FIRST-detected DOGs have compact radio emission on the scale of $\sim 5\arcsec$, at the angular resolution of the survey.

\section{Discussion}

Radio emission is clearly detected in the stacked images of the otherwise individually undetected PL DOG positions. We can derive the characteristic $1.4$-GHz radio power ($P_{1.4\text{\,GHz}}$) corresponding to this detection following the procedure described in \citet{Perger_stack}. The cumulative flux density can be given as
\begin{equation}
S_\text{sum}=\sum_{i=1}^{N} \frac{P_{1.4\text{\,GHz}}}{4\pi}\frac{1}{D_{\text{L,}i}^2\left(1+z_i\right)^{-\alpha-1}}
\end{equation}
where $D_{\text{L,}i}$ and $z_i$ are the luminosity distance and redshift of the $i$-th individual source, respectively, $N$ is the number of stacked PL DOG cutout images, and $\alpha$ is the radio spectral index defined according to the $S \propto \nu^{\alpha}$ convention where $S$ is the flux density and $\nu$ is the frequency. We used the PL DOG images of the \citetalias{DOGsample} sources, thus $N=224$. We added the values of the central $10\text{\,pixel}\times10\text{\,pixel}$ region of the summed cutout images, and obtained $S_\text{sum}=931.8$\,mJy. We assumed several values for the spectral index between $0$ and $-1.0$ and found that the resulting $P_{1.4\text{\,GHz}}$ always exceeds $3 \times 10^{22}\text{\,W\,Hz}^{-1}$. 

Compared to the extreme starburst galaxies of the local Universe, Arp\,299-A and Arp\,220, the characteristic radio power obtained for stacked PL DOGs exceeds their $1.7$-GHz radio power of $4.5 \times 10^{21}$\,W\,Hz$^{-1}$ and $1.6 \times 10^{22}$\,W\,Hz$^{-1}$, respectively \citep{Alexandroff2012}. On the other hand, even more prolific starburst galaxies with star formation rates as high as several $100\, \mathrm{M}_\odot\mathrm{\,yr}^{-1}$ can be found in the distant Universe \citep{bump_selection2}, closer to the star formation peak ($z\sim2$). According to \citetalias{DOGsample}, the estimated average infrared luminosity of their PL DOG sources is $\sim 10^{12.9} \mathrm{L}_\odot$. Using the relation of \cite{Kennicutt_1998}, this implies a star formation rate of $\sim1370\, \mathrm{M}_\odot\mathrm{\,yr}^{-1}$.

No similar significant radio emission is seen at the stacked B and U DOG positions. In principle, this could happen if the U and B DOG samples have higher redshifts on average, thus they have much fainter radio emission than the PL DOGs. However, \citetalias{DOGsample} investigated the redshift distribution of the PL and B DOG types and found that there is no significant difference between the redshift distribution of the two groups. We also calculated the averages and the standard deviations of the $z_\text{ph}$ values within the three groups of stacked DOGs (Table~\ref{tab:stacking}), and found practically no differences between them. Similarly, the difference in the detections of radio emission of the extended samples of PL and B DOGs cannot be ascribed to the different redshift distributions either (Table~\ref{tab:stacking}).

The difference in radio emission of the PL and B DOGs can also arise if the two groups have different luminosities. According to \citetalias{DOGsample}, the average infrared luminosity of the PL and B DOG sources are comparable, $L_\mathrm{PL}=(12.8\pm0.6)\,\mathrm{L}_\odot$ and $L_\mathrm{B}=(12.7\pm0.7)\,\mathrm{L}_\odot$. When comparing only those having reliable $z_\mathrm{phot}$ values, the B DOGs have slightly higher luminosity, $(13.0\pm0.3)\,\mathrm{L}_\odot$. However, they still agree within the errors with that of PL DOGs, ($12.9\pm0.2)\,\mathrm{L}_\odot$. The possible inaccuracy of the photometric redshift estimates (the DOG sources of the \citetalias{DOGsample} have only $z_\mathrm{phot}$) can result in uncertain luminosity values. Therefore, we compared the average $W4$ ($22\,\mu$m) flux densities of the stacked PL and B DOG samples as well. The weighted average $W4$ flux density of B DOGs, $\sim7.5\,\mu$Jy, is slightly lower than that of the PL DOGs, $\sim8\,\mu$Jy.

The larger infrared brightness of PL DOGs (compared to B DOGs) can be explained by slightly higher star formation rates, and can also cause the slightly enhanced radio emission found in PL DOGs. Alternatively, the higher infrared flux density can indicate the contribution from AGN in PL DOGs, in agreement with the explanation of their infrared SED. Generally, only $10$ per cent of AGN are radio-emitting \citep[e.g.][]{Ivezic2002}. Thus, even if we hypothesize that the radio emission of the stacked PL DOGs is due to AGN, we cannot expect that all of the stacked PL DOGs harbour a radio-emitting AGN. Therefore, the obtained characteristic $P_{1.4\text{\,GHz}}$ can be regarded as a composite value. In that scenario, the majority of the PL DOGs are expected to have moderate radio emission which originates from star formation only, but a smaller group of them, presumably $\sim10$\ per cent of the FIRST-undetected PL DOGs, do indeed harbour a radio-emitting active nucleus whose contribution to the total radio emission could increase the characteristic radio power derived by our stacking analysis.

\citetalias{DOGsample} identified DOG sources in their sample with excess blue colours which were similar to the optically selected BOSS quasars at similar redshifts. They selected these blue-excess DOGs based upon their optical spectrum. Assuming a power-law shape, these sources have an optical spectral index $<0.4$. None of these blue-excess DOG was detected in FIRST. All but one of them are PL DOGs. \cite{Assef2016} found similar objects with excess blue light among the extreme subgroup of DOGs, the hot DOGs. The blue-excess DOGs of \citetalias{DOGsample} do not satisfy the criteria of hot DOGs \citep[e.g.,][]{Eisenhardt2012}, however, \citetalias{DOGsample} explains the brighter blue emission following \cite{Assef2016} as leaked optical emission from the AGN, either scattered or direct light. To ascertain whether these obscured AGN are responsible for the radio emission revealed in the stacked images of PL DOGs, we repeated the stacking of PL DOGs with excluding the blue-excess DOGs from the sample. From the $7$ blue-excess PL DOGs \citepalias{DOGsample}, $4$ were included in our stacking procedure. If we exclude the $4$ FIRST cutout images corresponding to these sources, the peak SNR does not change in any noticeable way. Therefore, the overall radio emission detected in stacked PL DOGs cannot be ascribed to these few presumably intrinsically optically very bright but obscured AGN in the sample.

In the stacked median image of B DOGs, a radio-emitting feature appeared at the centre with $\text{SNR} \approx 4$. Our analysis of random positions showed that a peak of $4$ times the image noise level can appear accidentally. However, when stacking the B DOG sample, the very faint radio emission appeared at the centre of the image, right in the position where the emission would be expected, while in the case of stacking the cutouts at random positions, the radio feature could appear at any part of the median image. This hints on a faint radio emission in B DOGs in general, at a much lower level of flux densities than in PL DOGs.

The non-detection of radio emission in the median image of U DOGs can be explained by the intrinsically lower radio luminosity of the sources in this group. \citetalias{DOGsample} expect that most of the U DOGs are B DOGs, with a few additional low-luminosity PL DOGs, which is in agreement of the implication of our stacking result. Since it is expected that most of the U DOGs are B DOGs, to increase the number of objects and thus to reduce the noise level, we also stacked together the U and B DOG samples of \citetalias{DOGsample}. However, we could not detect radio emission, the SNR at the centre of the resulting image was below $3$. 

The FIRST-detected DOGs of the \citetalias{DOGsample} sample have higher $W4$ flux densities than the average values of stacked DOGs. The weighted average of FIRST-detected PL DOGs is $\sim 9.5\,\mu$Jy, and the $W4$ flux density of the one FIRST-detected B DOG is $(43.5\pm 2.5)\,\mu$Jy. Assuming a radio spectral index of $\alpha=0$, we calculated the $1.4$-GHz radio powers\footnote{Steeper radio spectra ($\alpha<0$) would result in larger radio power values.} for the FIRST-detected DOGs where a redshift estimate (either photometric or spectroscopic) is available. For one source of the \citetalias{DOGsample} sample, HSC\,J140638.20$+$010254.6, we found a spectroscopic redshift, $z=0.236$, in the literature \citep{SDSS_DR9}. We used that in the radio power calculation. The results are listed in Table~\ref{tab:FIRST_det}. Assuming that the radio emission solely originates from star formation, the star formation rate can be estimated from the $1.4-$GHz radio power values using the formula of \cite{Hopkins_SFR}. Except for the single detected B DOG (HSC\,J140444.81$+$002451.1) and one of the faintest PL DOG (HSC\,J140738.47$+$002731.4), the star formation rates all exceed $100 \mathrm{\,M}_\odot\mathrm{\,yr}^{-1}$, and in some cases, it even reaches $(10^3-10^4)\mathrm{\,M}_\odot\mathrm{\,yr}^{-1}$. On the other hand, if a radio-emitting AGN contributes to the detected radio power, the star formation rates can be lower. The existence of possible radio-emitting AGN in the FIRST-detected DOG sources can be verified via very long baseline radio interferometric (VLBI) observation. If an extragalactic radio source at $z\gtrsim0.1$ can be detected at mas-scale resolution with VLBI technique, the radio emission has to originate from an AGN \citep{Middelberg2013}.

\section{Conclusions}

We investigated the arcsec-scale radio emission of different DOG samples using the FIRST survey catalogue. We used the DOG samples compiled by \citetalias{DOGsample} employing Subaru HSC, VIKING DR2 and AllWISE data, and the DOG samples of  \citetalias{Bussmann2012} and \citetalias{Melbourne2012} based upon {\it Spitzer Space Telescope} measurements. \citetalias{DOGsample} classified the sources into three groups: PL, B, and U DOGs. The \citetalias{Bussmann2012} and \citetalias{Melbourne2012} DOGs are grouped into PL and B DOGs. 

Few DOGs were detected in the FIRST survey, most of them are PL DOGs. We stacked the FIRST undetected (i.e. with $1.4$-GHz radio flux densities below $\sim 1$\,mJy) DOGs and found significant radio emission in the median-stacked image of PL DOGs. This radio emission may originate from enhanced star formation compared to the other DOG sources and/or can be ascribed to the contribution of faint radio-emitting AGN possibly contained in some of the stacked PL DOGs.

There is no indication of radio emission in the stacked U DOG positions, even though a slightly lower noise limit than for the PL DOGs could be reached with the stacking analysis. In the case of the B DOGs, a marginal detection with $4$ times the image noise level can be seen in the median-stacked image. However, stacking FIRST image cutouts from random positions showed that a similar number of cutouts can give rise to peaks at any location within the field in the median-stacked image with comparable SNR. Still, in the case of B DOGs, the $4\sigma$ radio emission appeared at the centre of the median image, hinting its relation to the stacked DOG sources. To unambiguously quantify the faint radio emission from B DOGs as a class of objects using a similar stacking procedure, a larger sample of source positions would be needed. 

With respect to the FIRST-detected DOGs, assuming that the 1.4-GHz radio power solely originates from star formation, we obtained star formation rates ranging from several hundreds to $10^4\mathrm{\,M}_\odot\mathrm{\,yr}^{-1}$. High-resolution VLBI observations can reveal whether radio-emitting AGN contribute to these high radio power values.

\section*{Acknowledgements}

We thank the anonymous referee for valuable comments which helped to improve the manuscript. This work was supported by the Hungarian National Research, Development and Innovation Office (OTKA K134213). K\'EG was supported by the J\'anos Bolyai Research Scholarship of the Hungarian Academy of Sciences.

\section*{Data Availability}

The datasets underlying this article were derived from sources in the public domain: FIRST survey database, \url{http://sundog.stsci.edu/}.



\bibliographystyle{mnras}
\bibliography{refDOG} 








\bsp	
\label{lastpage}
\end{document}